\documentclass{aa}
\usepackage{graphics}

\def\etal {et al.~}

\def\ltsima{$\; \buildrel < \over \sim \;$}
\def\simlt{\lower.5ex\hbox{\ltsima}}
\def\gtsima{$\; \buildrel > \over \sim \;$}
\def\simgt{\lower.5ex\hbox{\gtsima}}

\def\vs{{~vs~}}

\def\kms{\ifmmode {\rm \ km \ s^{-1}}  
\else
$\rm km \ s^{-1}$\fi}

\def\h2{\mbox{\ion{H}{ii}}}

\newcommand{\lessim}{\mathrel{\hbox{\rlap{\hbox{\lower4pt\hbox{$\sim$}}}\hbox{$<$}}}}

\begin{document}
\addtolength{\voffset}{1cm}

\thesaurus{10.19.3; 08.19.1}

\title{Constraints on Thick Disc and Halo parameters from HST photometry of 
field stars in the Galaxy}

\author{L. O. Kerber, S. C. Javiel, B. X. Santiago}

\offprints{kerber@if.ufrgs.br}
 
\institute{Universidade Federal do Rio Grande do Sul, IF, 
CP\,15051, Porto Alegre 91501--970, RS, Brazil}

\date{Received 11 July 2000 / Accepted 19 October 2000}

\titlerunning{Constraints on Thick Disc and Halo parameters}
\authorrunning{Kerber, Javiel \& Santiago}

\maketitle

\begin{abstract}
 We analyze a sample of over 1000 stars from 32 fields imaged in the V and I
bands with the {\it
Wide Field and Planetary Camera}, on board of the Hubble Space Telescope.
The fields are located at Galactic latitudes $\vert b \vert \geq 15^{\circ}$
and in various directions on the sky. We consider models for the structure
of the Galaxy with different choices for the main parameters governing the
shape and luminosity function of the thick disk and stellar halo. Comparing
model predictions with the observed colour-magnitude diagram we are able to
rule out an increasing or flat stellar luminosity function 
at the low-luminosity
end. We also rule out large values of the vertical scale height of the
thick disc, $z_0$, finding it to be in the range $800 \leq z_0 \leq 1200$ 
pc.
As for the local density normalization, values within the range $4 \% \leq 
n_0 \leq 8 \%$ seem to better reproduce
the data. Our data essentially rule out a flattened stellar halo ($c/a$ \ltsima
$0.5$) or models with both large local normalization and effective radii.


\end{abstract}

\section{Introduction}
\label{intro}
Our Milky Way galaxy is known to have a large disk with spiral structure,
typical of luminous galaxies of Sb type. It has also been known for a long 
time to 
contain a population of old ($\tau$ \gtsima $10$ Gyrs) stars distributed in
an spheroidal component. More recent models 
of its structure have included at least
a third component, of intermediate properties, often called the 
{\it thick disc } (Gilmore \& Reid 1983). Our 
understanding of the
spatial, kinematical and chemical structure of the Galaxy has continuously
improved in more recent years as a result from the quest for 
a more quantitative
and detailed picture of how the Galaxy (and galaxies in general) formed and
evolved. However, despite the enlarged and improved stellar samples 
with accurate photometric, parallax and velocity measurements, 
several issues concerning the structure, 
origin and evolution of the main components and populations in the
Galaxy still remain to be settled (Gilmore \etal 1995, Majewski \etal 1996,
Norris 1999).

The very existence of the thick disc as a discrete component, 
with stars kinematically and chemically distinct from those of 
the thin disk and halo, is subject to controversy (Norris 1987, 
Carney \etal 1989, Reid 1998). There is little doubt, however, 
that models with one planar component, with a single density 
profile, do not provide a suitable description of the structure 
of the Galaxy, since such models do not successfully fit recent 
star count data (Santiago \etal 1996a, Buser \etal 1999).

The structure of the thick disc is generally described as a double exponential,
with horizontal and vertical scales in the ranges 
$r_0 \simeq 2.5-4.0$ kpc and $600$ pc \ltsima $z_0$ \ltsima $1600$ pc, 
respectively
(Chen 1996, Robin \etal 1996, Buser \etal 1999, 
Norris 1999). 
The wide range of scale height values reflects in part the
difficult task of defining and separating thick disc stars from those of other 
structural components. The value of $z_0$ for the thick disc 
is anti-correlated with the 
local density of intermediate population stars, 
$n_{0D}$, which normalizes the assumed
density profile: models with larger $z_0$ usually have smaller $n_{0D}$ values
and vice versa.

The spatial structure, stellar luminosity distribution and shape of the 
outer regions of the spheroid, the stellar halo, have also been under closer
scrutiny in recent years (Wetterer \& McGraw 1996, Gould \etal 1998, 
Stetson \etal 1999, Moore \etal 1999, Samurovic \etal 1999). 
Largely due to Hubble Space Telescope (HST)
observations, it is now known that the stellar luminosity function (LF) both of
globular clusters and field disk stars decreases at the low-luminosity end
(De Marchi \& Paresce 1995a,b; Elson \etal 1995; Santiago \etal 1996a,b; 
Gould \etal 1996,1997, Piotto \etal 1996, 1997, Mendez \& Guzman 1998). 
However, there is no consensus yet about the shape of the 
stellar LF at the faint end for other components (Dahn \etal 1995,  
Gould \etal 1998). Also, the flatenning 
of the stellar halo and the density profile function that best describes it
are subject to uncertainty.

Improved observational data will more efficiently constrain the models if
use is made of objective and statistically sound means of comparing 
model predictions to observations. Recent efforts in the direction of
efficiently modeling observed colour-magnitude diagrams
(CMDs) or other N-dimensional spaces of observables 
have been developed (Saha 1998, Hernandez \etal 1999, 
Lastennet \& Valls-Gabaud 1999, Stetson \etal 1999). These
methods usually try to make use of non-parametric statistics and a minimum
amount of initial assumptions. 

In this paper we analyze a composite CMD of 32 stellar
fields towards various directions 
in the Galaxy, all imaged with HST's WFPC2. We apply simple and
objective statistical tools to compare
the observed CMD with those generated using structural models of the Galaxy. 
In Sect. 2 we discuss the data used, whereas
in Sect. 3 we present the statistical methods and models. In Sect. 4 we discuss
the models found to best reproduce the data 
and the consistency among the different statistics used.
Finally, in Sect. 5 we present our final conclusions and future perspectives.

\section{The WFPC2 data}
Our WFPC2 fields are part of the Medium Deep Survey (MDS) database.
The data were extracted from 32 deep fields, 17 of which, 
mostly at high Galactic
latitudes, were analyzed by Santiago \etal (1996a). 
The new 15 fields are listed
in table 1. Their MDS id is given, along with their Galactic longitude and
latitude ($l_{II}$ and $b_{II}$, respectively), total number of exposures
and exposure times and the number of stars found in each. 
All fields have
been observed with the $F606W$ and $F814W$ filters and have at least 2 
exposures per filter. Several of the 15 new fields have $\vert b_{II} \vert$ 
values in the range  $[15^{\circ},30^{\circ}]$, thus containing a larger fraction
of thick disc stars than those studied by Santiago \etal (1996a).

\begin{table}
\caption[]{The new 15 fields.}
\label{tab1}
\small
\renewcommand{\tabcolsep}{1.1mm}
\begin{tabular}{lrrrrrrr}
\hline\hline
{Field} & {$l_{II}(^\circ)$} & {$b_{II}(^\circ)$} & {$t_{\it F814W}$} & {$t_{\it
F555W}$} & {$\# I$}   & {$\# V$} & {$\# stars$} \\ \hline
\hline
u6w0 &  273.691 & -15.841 &  4400 &  1200 & 4 & 2 &  114 \cr
udm1 &  172.759 & -51.358 &  5400 &  4000 & 3 & 4 &   14 \cr
umdc &  81.826 & -19.165 & 4200 &  2400 & 2 & 2 &   52  \cr
uo50 &  43.671 &  20.340 & 6300 &  3300 &  3 & 2 &  212 \cr
uop0 & 206.073 &  19.625 & 4200 & 7200 & 2 & 5 &     36 \cr
uqc01 & 73.076 & 26.259 &  3600 & 3900  & 2 & 2 &    46 \cr
urw10 & 279.058 & 33.584 & 1000 & 1250 & 2 & 2 &     30 \cr
uwy02 & 299.632 & 51.019 & 9600 & 11700 & 5 & 6 &    32 \cr
uzp0  & 202.267 & 76.454 & 12600 & 6600 & 6 & 4 &    12 \cr
ueh0 &  123.681 & -50.299 & 12600 & 8700 & 6 & 5 &    8 \cr
uem0 &  178.475 & -48.117 & 6600 & 2400 & 5 & 2 &    12 \cr
ust0 &  247.868 &  36.895 & 23100 & 16500 & 11 & 10 & 9 \cr
uui0 &  130.508 &  44.473 & 6300 & 5400 & 3 & 3 &     0 \cr
uy00 &  359.019 &  64.701 & 6900 & 6000 & 4 & 4 &    13 \cr
uzk0 &  154.610 &  75.121 & 11100 & 8400 & 5 & 5 &    4 \cr
\hline\hline
\end{tabular}
\end{table}

\begin{figure} 
\resizebox{!}{\hsize}{\includegraphics{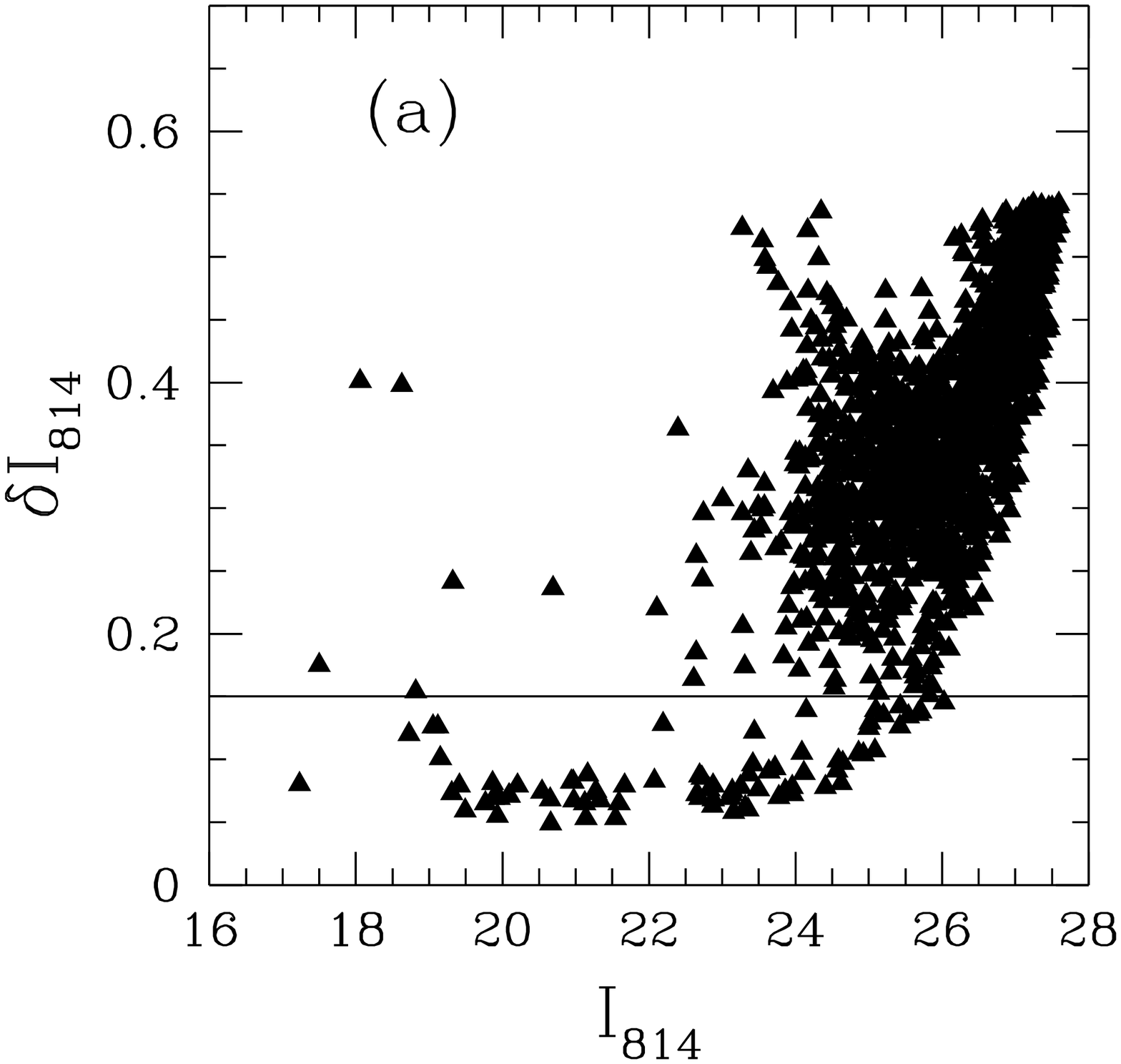}}
\resizebox{\hsize}{!}{\includegraphics{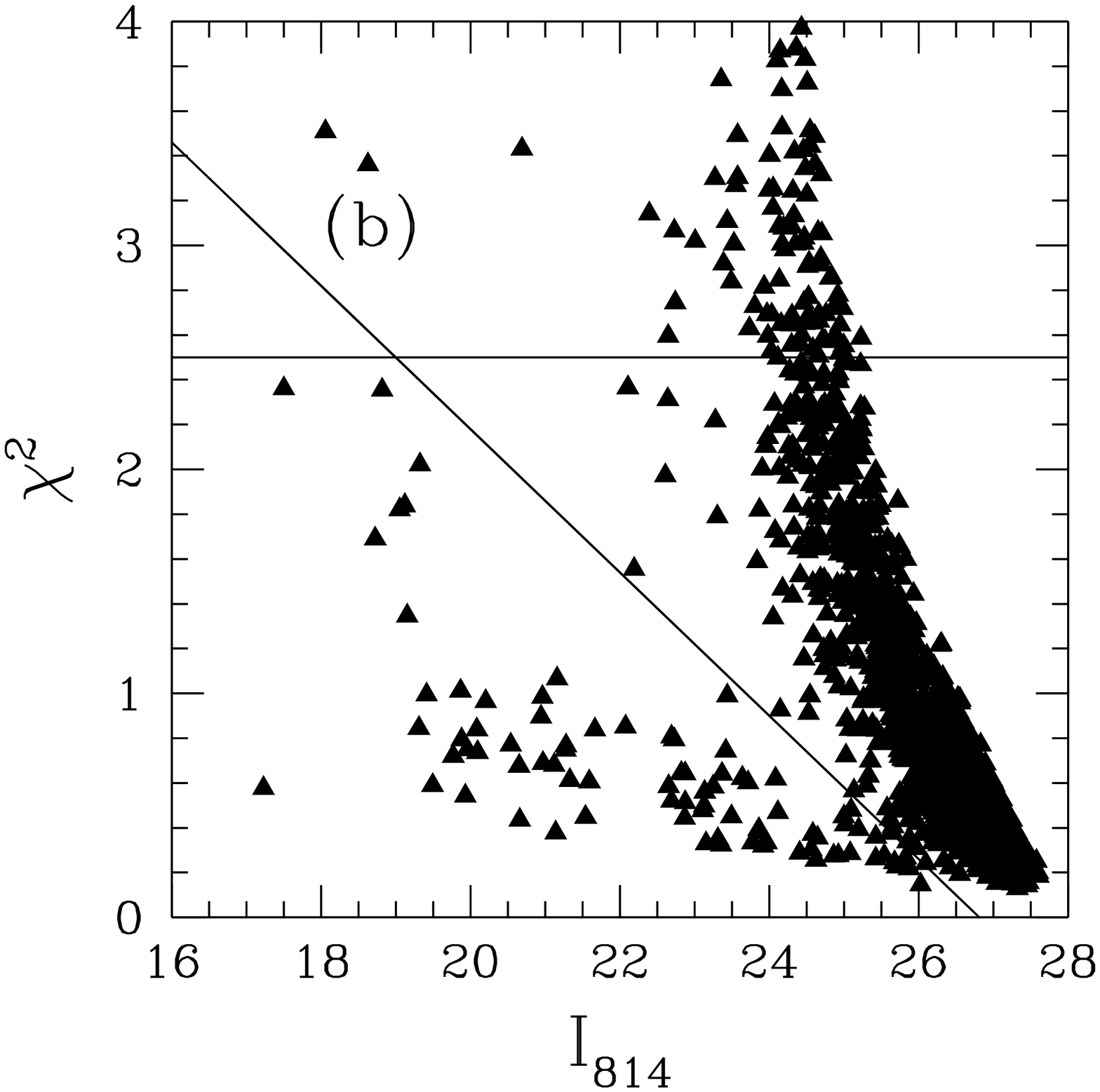}}
\caption[]{{\it a)} $\delta m \vs I_{F814}$ plot for one of our
intermediate latitude fields. {\it b)} $\chi^2 \vs I_{F814}$ plot
for the same stars on panel {\it a}. Indicated in both panels are the
cuts done in order to separate real stars from extended or spurious objects.}
\label{figuo50}
\end{figure}

Sample definition, photometry, instrumental and completeness corrections,
and calibration to the standard system for the new fields 
were carried out in a similar fashion
as in Santiago \etal (1996a). In brief, the DAOPHOT package within IRAF
was used to automatically make an object list (whose peak intensities were
$5\sigma$ above sky background) and to measure magnitudes both using aperture 
photometry and point spread function (psf) modeling. A single psf template
was built using a compilation of very high signal-to-noise isolated stars
from the different fields. Star/galaxy separation proceeded by applying
cuts in the space of parameters output by the psf fitting task ALLSTAR.
Fig. 1 shows the cuts applied to the $\delta m \vs I_{814}$ (panel {\it a})
and $\chi^2 \vs I_{814}$ (panel {\it b}) relations in the field uo50. 
From these panels
we also determined cut-off magnitudes both in the bright end
(due to saturation, where $\chi^2$ sharply increases) and in the faint
end (where disentangling stars from other sources becomes impossible).
For the fields with too few stars for that to be done, we used
the cut-off magnitudes from the more crowded fields, 
scaled up or down according to exposure time.

Photometric corrections for limited aperture and charge-transfer effect (CTE),
as prescribed by Holtzman \etal (1995a,b) were applied to the data. 
The magnitudes and colours were converted to the standard system, also 
using the transformations listed by those authors. Finally, completeness
levels were obtained as a function of the $F814W$ magnitudes and used to
compensate for the loss of stars in the faintest magnitude bins.
Typically, completeness started to drop abruptly from 100\% at $I_{814W} 
\simeq 24$

In Fig. 2 we show the joint CMD for all 32 WFPC2 fields. 
It contains over 1000
stars distributed within a large range of magnitudes and colours. The 
magnitudes and colours were obtained by aperture photometry,
are in the standard photometric system and corrected
for the several effects mentioned above. Notice, however, that no correction
to redenning was applied to the data. We preferred to incorporate 
redenning effects to our 
model predictions, described in the next section. 

\begin{figure} 
\resizebox{\hsize}{!}{\includegraphics{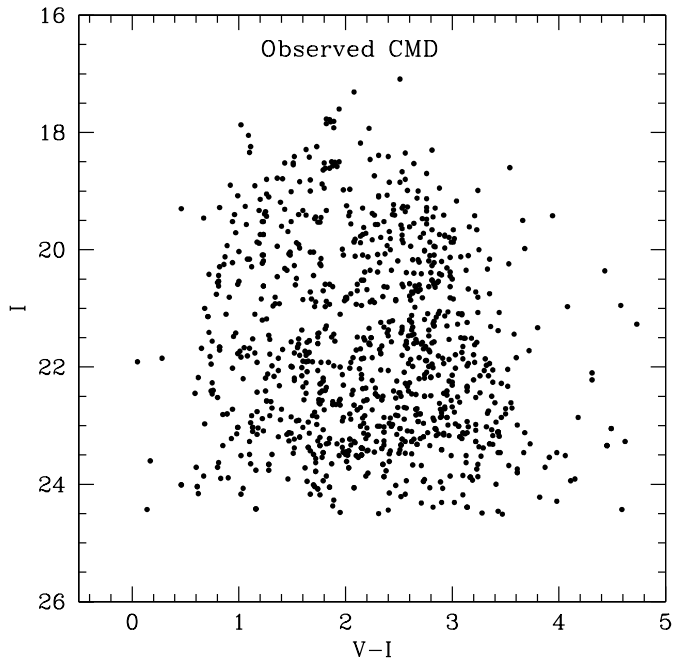}}
\caption[]{The observed composite CMD for all 32 fields included in our
work. The magnitudes and colours are corrected for instrumental effects
(aperture and CTE) and calibrated to the standard system.}
\label{figobs}
\end{figure}

\section{Models for the Galaxy and Statistical Techniques}

\subsection{Modeling the Galaxy}

Our main goal is to compare the observed CMD of HST/WFPC2 stars with 
theoretical CMDs based on models for the structure of the Galaxy. The model
number counts as a function of colour and magnitude can be obtained from
numerical integration of the Fundamental Equation of Stellar Statistics.
The main ingredients entering this equation are the stellar luminosity
function, $\Phi (M)$ (LF), and the density distribution $\rho ({\bf r})$ 
of stars for each structural component in the model. We consider models 
with 3 major
components: a thin disc and a thick disc, both with double exponential 
density profiles, and a spheroidal or halo component, 
with a deprojected {\it de Vaucouleurs}
profile. The main parameters governing these profiles are the horizontal 
and vertical exponential scales, $r_0$ and $z_0$ respectively, and the
effective de Vaucouleurs radius, $R_e$, for the halo. 
Also relevant are the normalization
factors, $n_0$, for the different components, which are usually 
expressed as percentage values
of the local spatial density of stars. 

The LF used in our modeling 
is the one originally from Wielen \etal (1983) up to $M_V = 12$. For lower
luminosities we treat it as a power law in luminosity, 
$log \Phi (M_V)~ \propto~ \alpha~M_V$
with slope $\alpha$ considered as a free 
parameter. In our data $\vs$ model comparisons (see below), we also allow for 
variations in $z_0$, $r_0$, $R_e$ and in the normalization of the
thick disc ($n_{0D}$) and spheroidal components ($n_{0H}$).
For more on the models, profiles and LFs
we refer to Reid \& Majewski (1993), Majewski \etal (1996) 
and Santiago \etal (1996a).

The numerical integration of stellar statistics was carried out for all 32 
fields, taking into account their variable available magnitude ranges (set 
by saturation and detection) and completeness 
functions. Galactic extinction was also incorporated to the models 
using the Burstein \& Heiles (1982) E(B-V) maps. 
Conversion of extinction vectors and
model predictions from 
BV to VI was carried out in the same way as in Santiago \etal (1996a).
The output from these integrals are tables with $(V-I)$ colour and $I$ 
magnitude counts for each field which can then be added together 
to yield a composite model CMD.

We explored an extensive grid of models, covering large regions of the
parameter spaces formed by the thick disc and halo structural 
parameters and LF slopes. 
Thick disc scale height and normalization values varied within the ranges
$500 \leq z_0 \leq 2000$ pc and $1 \% \leq n_{0D} \leq 8 \%$, respectively.
But notice that these two parameters were paired so that their anti-correlation
was respected. A total of 32 thick disc models resulted. For each one of
them, we considered 4 choices of the horizontal thin disk
e-folding length: $r_0 = 2500, 3000, 3500, 4000$ pc, and 2 choices for the
LF slope at its low-luminosity end ($M_V \geq 12$): $\alpha = -0.2$ 
and $\alpha = 0$, totalling 256 thick disc models.

As for the halo we used $R_e = 1500, 3000, 4500$ pc for its
effective radius and $n_{0H} = 0.1, 0.125$ and $0.2\%$ for the
local normalization. Another halo parameter varied was its axial ratio, with
values $c/a = 0.5, 0.65, 0.8, 1.0$, yielding a total of 36 stellar halo
models.

This extensive parameters grid was defined so as to include the variety 
of parameter values often
quoted in the literature (Norris 1999) and to accommodate most thick
disc and halo parameters whose values are still uncertain.

\subsection{Statistical Tools and Techniques}

Fig. 3 shows a comparison of a theoretical CMD, whose main parameters are 
indicated, with the observed one. The simulated CMD was built from the
expected model number counts as a function of colour and magnitude using
Monte-Carlo techniques: we randomly threw points on the CMD plane with
probabilities proportional to the expected number in each CMD position.
The total number of points in the simulated CMD was a Poisson deviate of the 
total model counts.

\begin{figure} 
\resizebox{\hsize}{!}{\includegraphics{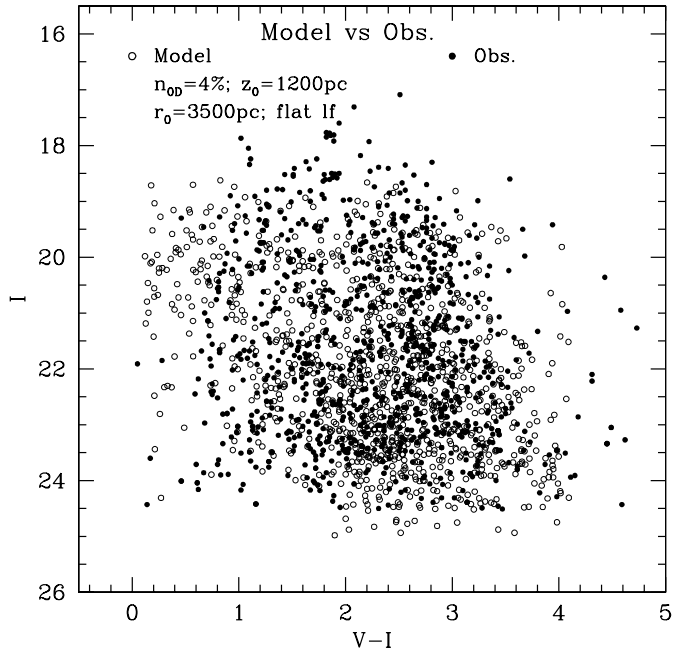}}
\caption[]{A CMD for a model realization compared to the data CMD.}
\label{figcmd}
\end{figure}

Our main challenge is to find an objective and
efficient statistical tool to quantify how different or similar 
these two two-dimensional distributions of points are. 
A visual comparison reveals regions in the CMD,
such as $0 \leq V-I \leq 0.8$ and $18 \leq I \leq 22$, where there is a clear
excess of model points. There are other regions 
(like $0 \leq V-I \leq 0.8$ and $23.5 \leq I \leq 24.5$, for instance) 
with more observed than model stars. 
The different models in the grid show a smooth and continuous
variation in their CMDs, making it hard to visually select best fitting models.

We here propose two statistics which are objective and simple tools for 
comparing two distributions of points on a plane. The first is simply a measure
of the dispersion between number counts within different data bins.
Let $m_i$ and $o_i$ be the number of model and observed stars 
in the $i^{th}$ of $N$ bins regularly spaced on the CMD plane; 
we then define the dispersion as:

$$s^2 = \sum_{i=1}^N(m_i - o_i)^2$$

The second statistics was used by Santiago \& Strauss (1992) in a different
context, to study spatial segregation of galaxies. It also compares the
number of CMD points found in different bins. Let $p_i$
be the percentile position of $\vert o_i - m_i \vert$ 
within the distribution of $\vert m_i - m_{ij} \vert$ 
($j = 1,N_{real}$) values, 
where $m_{ij}$ are Poisson realizations of the expected 
model counts $m_i$ in the $i^{th}$ bin. We then define

$$pss =\sum_{i=1}^N log (100 - p_i (\%) )$$

We typically compute $pss$ from a distribution of $N_{real} = 300$ 
Poisson realizations of each model.
Even though both statistics depend on data binning, we have observed that both
are quite insensitive to the binning scheme. They are also applicable to
datasets with varying number of points. 
This is in contrast with what we found for the W statistics (Saha 1998),  
whose results seem to be sensitive to the total number of points in both 
distributions being compared.

\begin{figure} 
\resizebox{\hsize}{!}{\includegraphics{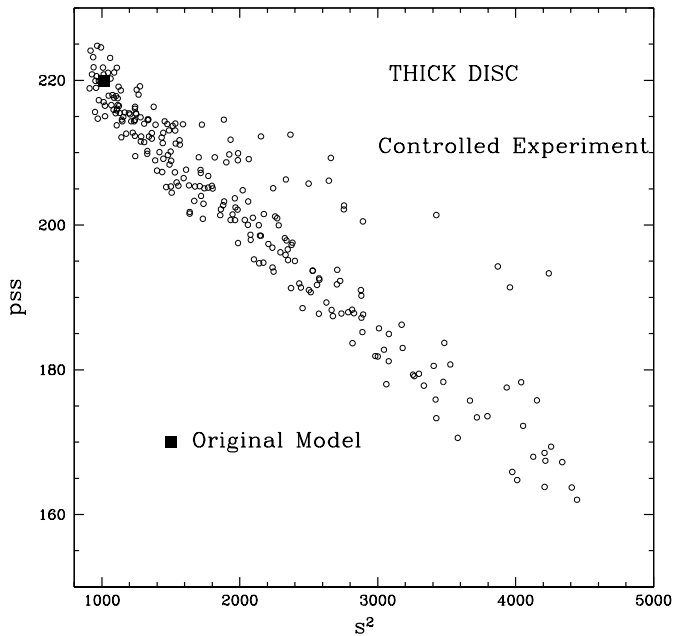}}
\caption[]{$Pss \vs s^2$ diagram for a controlled experiment. Each point
corresponds to a model with a particular set of values for the thick disc 
structure and LF. The large dot gives the position of the model from which the
``data'' CMD was created.}
\label{figexp}
\end{figure}

Clearly, the better the agreement between model and observed CMDs, the smaller
$s^2$ and the larger $pss$ will be. We tested both statistics by means of
controlled experiments, in which we generated a Monte Carlo realization from a 
particular model (as described in the beginning of this section) 
and compared it to all the models. Fig. 4 shows the result
of a typical such experiment. It shows a diagnostic diagram of $pss \vs s^2$
values obtained by comparing the simulated ``data'' with the theoretical
CMD for all 256 models in the thick disc grid described in the previous 
section. There is a clear anti-correlation between the 2 statistics, as 
expected.

The large dot in Fig. 4 corresponds to the particular model used to generate
the ``data'' points. It is clearly among the extreme values of both statistics,
again as expected. However, the fact that it is not the most extreme 
point in the
upper left of Fig. 4 is a measure of the resolution provided by the 
diagnostic diagram used.

\section{Results}
\subsection{Thick Disc}

In this section we apply the $pss$ and $s^2$ statistics to 
the comparison between
our observed CMD and the theoretical CMDs from the grid of models presented
earlier. Our main tool is the $pss \vs s^2$ diagram. In computing $pss$
we ran $N_{real} = 300$ Poisson realizations for each model in order 
to build the 
distribution of $m_{ij}, j=1,300$ model number counts in each bin. 
We adopted a 14x10 binning covering
the I \vs (V-I) plane in the ranges $18 \leq I \leq 25$ and 
$0 \leq (V-I) \leq 4$.

Fig. 5 shows the position of all 256 models we considered by varying the
thick disc structural parameters $z_0$, $n_{0D}$ and $r_0$ and its stellar
LF slope. There is again a clear correlation among the two statistics. 
Furthermore, the position within this diagram is correlated with the parameter
values. The diagram splits into two branches, the upper one corresponding to
models with 
$z_0 \geq 1200$ pc, which usually feature a larger than observed 
number of stars.
The lower branch (small pss and $s^2$ values) contains models with a deficiency
of stars relative to the observed sample. The best models in the upper left
have $800 \leq z_0 \leq 1200$ pc and
$4 \leq n_{0D} \leq 8\%$, with little or no dependence on the horizontal
exponential scale, $r_0$.  

\begin{figure} 
\resizebox{\hsize}{!}{\includegraphics{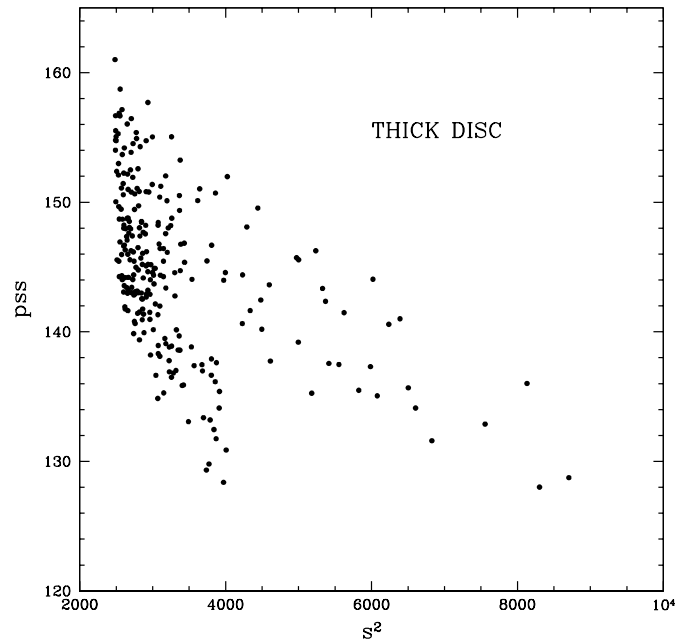}}
\caption[]{The $pss \vs s^2$ diagram obtained by comparing 256
thick disc models  to the real data.}
\label{figth}
\end{figure}

\begin{figure} 
\resizebox{\hsize}{!}{\includegraphics{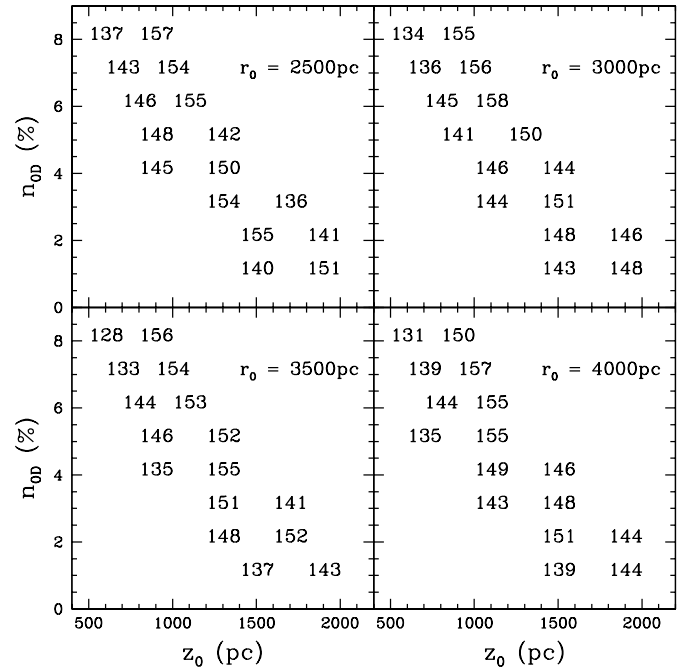}}
\caption[]{$Pss$ maps in the $z_0 \vs n_{0D}$ grid for 4 different values
of the horizontal scale length $r_0$, as indicated.}
\label{figmap}
\end{figure}

These results are further confirmed by Fig. 6, where
the $pss$ values (assuming a decreasing LF at the faint end) are shown 
for different locations within the
$z_0 \vs n_{0D}$ plane for fixed $r_0$.
The diagrams do not differ significantly, with the best $pss$ values 
located within the range $800 \leq z_0 \leq 1200$ pc and
$4 \leq n_{0D} \leq 8\%$ for all panels. Note that the anti-correlation between
these two structural parameters persists, a $z_0 \simeq 800$ pc with
$n_{0D} \simeq 8\%$ being essentially as good a model as one with $z_0 \simeq
1200$ pc and $n_{0D} \simeq 4\%$.

There is also a clear trend for models with decreasing faint end LF slope
to be favoured by our observed CMD. 
In Fig. 7 we show the difference in $pss$ values 
between models with decreasing
and flat LF slopes at the faint end, all the other parameters being the
same. The large majority of the 128 such differences are positive, 
indicating that the data are best described by a 
decreasing LF. In fact, of the 35 models located closer to the upper left
boundary of the $pss \vs s^2$ diagram, only 1 had a flat thick disc LF slope.
A similar result of a decreasing faint end for the LF has also been found
for the halo and/or disk stars by
Bahcall \etal (1994), Santiago \etal (1996a), Gould \etal (1997) 
and Mendez \& Guzman (1998) using different data sets.

\begin{figure} 
\resizebox{\hsize}{!}{\includegraphics{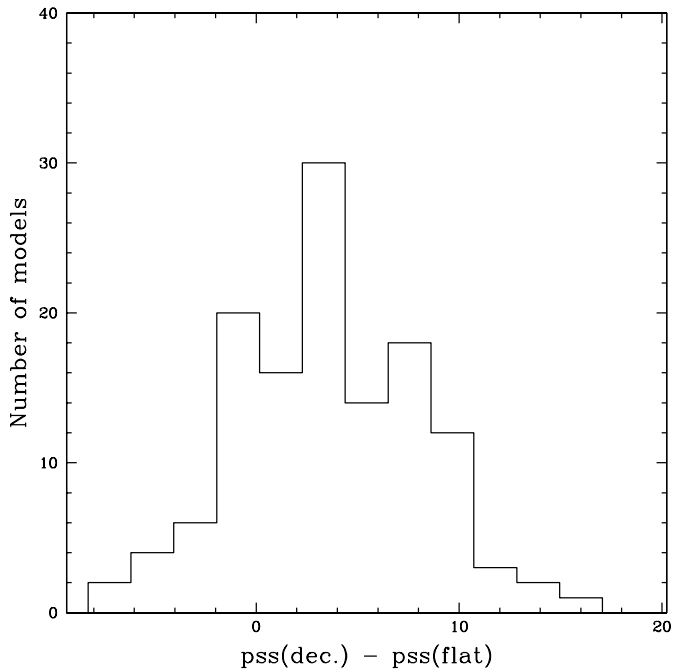}}
\caption[]{Distribution of $\Delta pss$ values for pairs
of thick disc models whose only difference is the value of $\alpha$, the
low-luminosity slope of the LF.}
\label{figlf}
\end{figure}

The consistency of the results just shown was tested by adopting both coarser 
(70 bins) and narrower (280 bins) binning schemes, as well as 
different color ranges than the one quoted early in this
section. There is no significant change in the $pss \vs s^2$ diagram or in the
position within it of models with different parameters. 

\subsection{\normalsize{Halo}}

In order to increase the fraction of halo stars in our observed CMD,
we have considered only the 18 WFPC2 fields with 
$\vert b \vert \geq 40^{\circ}$ and which point away from the centre of the 
Galaxy. This reduced, higher latitude, sample contains over 200 stars.
Their CMD was then compared to those generated by the 36 halo models discussed
at the end of Sect. 3.1. 

\begin{figure} 
\resizebox{\hsize}{!}{\includegraphics{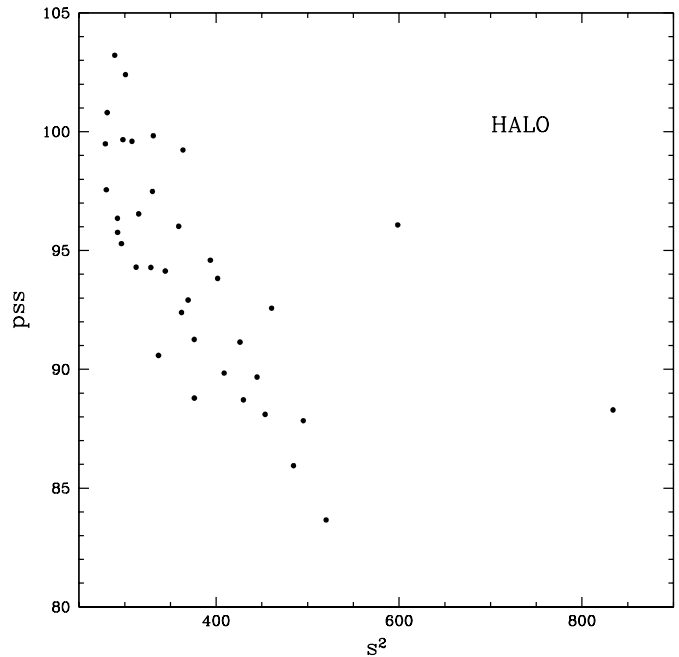}}
\caption[]{The $pss \vs s^2$ diagram obtained by comparing 36
halo models to the real data.}
\label{fighalo}
\end{figure}

The $pss \vs s^2$ diagram for these 36 models is shown in Fig. 8.
We again notice the anti-correlation between the two statistics.
And again there is a correlation between position within this diagram and
model parameters. The 2 models above the main branch have
$n_{0H} = 0.2\%, c/a=1.0$ and large values of $R_e$ ($R_e =3000$ or $4500$ {bf pc}). 
Most importantly, models with $c/a = 0.5$ are concentrated towards 
the lower part of the diagram, therefore being ruled out. Due 
to the smallness of the sample, however, it was not possible to single
out any other region of parameter space favoured by the data.

\section{Summary and Conclusions}

We carried out V and I photometry for over 1000 field stars in the Galaxy
down to faint ($I \simeq 25$) magnitudes and distributed over different
directions on the sky imaged with deep HST/WFPC2 exposures. The CMD for 
the data, corrected for instrumental effects and calibrated, was compared with
theoretical CMDs expected from a grid of models for the structure of the 
Milky-Way. The main grid parameters are the thick disc scale height $z_0$,
its local density normalization, $n_{0D}$, the stellar halo effective
radius $R_e$, normalization $n_{0H}$ and axial ratio $c/a$, as well as the
slope of the luminosity function at its low-luminosity end ($M_V \geq 12$).

We used two simple statistical tools to compare model and observed
CMDs, in an attempt to restrict the range of acceptable values for the
structural and LF parameters of the halo and thick disc.
Our analysis indicates acceptable models in the range
$800 \leq z_0 \leq 1200$ pc, $4 \leq n_{0D} \leq 8\%$ with these two parameters
being anti-correlated.
These ranges in both parameters are substantially narrower than those
quoted in a recent review by Norris (1999). In particular, our results do not
accommodate the very thick disc of $M > 0.3 M_\odot$ stars suggested 
by Gyuk \& Gates (1999) in their reinterpretation of microlensing results.
The $z_0$ range inferred from our work is also in disagreement 
with earlier works using ground-based star counts or samples of 
RR Lyrae stars, which have yielded $z_0 \geq 1500~pc$ 
(Hartwick 1987, Rodgers 1991, Reid \& Majewski 1993). On the other 
hand, as pointed out by Majewski (1993), introduction of metallicity 
gradients in our models could yield larger $z_0$ values. Another 
possibility, along similar lines, is that the thick disc may have a 
complex structure with possible subcomponents, such as the metal-weak 
thick disc (Chiba \& Yoshii 1998, Norris 1999).
Finally, our results are only marginally
consistent with that of Robin \etal (1996), who find $z_0 = 760 \pm 50$ pc
and $n_{0D} = 5.6 \pm 1 \%$. A better agreement is found with 
Buser \etal (1999), whose best model for the thick disc has
$n_{0D} = 5.9 \pm 3\%$ and $z_0 = 910 \pm 300$ pc.

As already established for disk stars by Santiago \etal (1996a), Gould \etal
(1996,1997) and
Mendez \& Guzman (1998) (from studies of HST/WFPC2 fields), 
we find the thick disc LF to be decreasing beyond $M_V
\simeq 12$ with large confidence. This is also in agreement with most
studies of globular cluster luminosity functions.

We also attempted to restrict the structural parameters of the halo using a
smaller sample with $\vert b \vert > 40^{\circ}$. We were able to rule out
flattened oblate halo models, with $c/a$ \ltsima $0.5$, or models
with both high normalization and high effective radius. The smallness
of the sample, however, prevents further conclusions about halo structure. 

A way forward will be to devise more and better statistical tools to be
applied to larger stellar samples. Particularly useful in constraining
the structure of the main components of the Galaxy will be the inclusion
of future kinematic data to be obtained from new orbital 
astrometric missions such as SIM and GAIA (Majewski 2000).

\begin{acknowledgements}
We are very grateful to David Valls-Gabaud for useful discussions and 
suggestions. We also thank Gerry Gilmore and Jim Lewis
for the original software used for modeling the structure of the Galaxy.
We also acknowledge CNPq and PRONEX/FINEP\,
76.97.1003.00 for partially supporting this work. 

\end{acknowledgements}

%
%

\end{document}